\documentclass[sigconf,nonacm]{acmart}
\usepackage{multirow} 
\setcopyright{none}
\usepackage{booktabs}
\usepackage{xcolor}
\usepackage{makecell}
\usepackage{subcaption}
\acmConference[KDD Cup 2026 UniRec Workshop]
{KDD Cup 2026 Tencent UniRec Challenge Workshop}
{August 12, 2026}
{Jeju, Korea}

\acmYear{2026}
\acmDOI{}
\acmISBN{}
\AtBeginDocument{%
  }

\usepackage[table]{xcolor}
\usepackage{multirow}
\usepackage{booktabs}
\newcommand\blfootnote[1]{%
  \begingroup
  \renewcommand\thefootnote{}%
  \footnotetext{#1}%
  \addtocounter{footnote}{-1}%
  \endgroup
}

\usepackage{hyperref}

\begin{document}

\title{From Feature Interaction to Feature Transport - A Unified Block for Scalable Recommendation Models}





\author{Zichen Luo}
\correspondingauthor
\affiliation{%
  \institution{Tianjin University}
  \city{Jinnan Qu}
  \state{Tianjin Shi}
  \country{China}}
  \email{luozichen@tju.edu.cn}

\author{Jiachen Guo}
\affiliation{%
  \institution{Tianjin University}
  \city{Jinnan Qu}
  \state{Tianjin Shi}
  \country{China}}

\author{Keming Gu}
\affiliation{%
  \institution{Tianjin University}
  \city{Jinnan Qu}
  \state{Tianjin Shi}
  \country{China}}

\author{Jie Zhang}
\affiliation{%
  \institution{Tianjin University}
  \city{Jinnan Qu}
  \state{Tianjin Shi}
  \country{China}}





\renewcommand{\shortauthors}{Luo et al.}

\begin{abstract}
Unified recommendation models aim to jointly model non-sequential multi-field features and sequential user behaviors, but existing interaction-centric designs mainly focus on mixing heterogeneous tokens within each layer. We argue that scalable unified recommendation also requires controlling how intent information is carried, filtered, and preserved across stacked blocks. Inspired by flow-based representation dynamics, we introduce feature transport, a view that treats deep unified recommendation as a discrete context-conditioned representation evolution process. We propose CRAFT, a Contextual Residual Adaptive Feature Transport block, which summarizes non-sequential features into a reliability-aware contextual field and uses it to generate residual displacement and memory-preserving signals for intent and sequence representations. In this way, non-sequential context acts as an active controller of representation evolution rather than a passive object of interaction. In the TAAC2026 advertising recommendation competition, CRAFT achieves a test AUC of 0.838090, surpassing the previous leaderboard-best score of 0.83798. Scaling experiments further show that CRAFT benefits from both depth and width expansion: stacking CRAFT to six blocks improves test AUC to \textbf{0.838148}, while increasing the hidden dimension reaches 0.838106. These results demonstrate the effectiveness, scalability, and generalization potential of the feature transport paradigm. Source code: https://github.com/AshleyLuo001/CRAFT
\end{abstract}
\maketitle
\blfootnote{\scriptsize
KDD Cup 2026 Tencent UniRec Challenge Workshop, August 12, 2026, Jeju, Korea.\\
Competition website: \url{https://algo.qq.com/}. em
}
\newcommand{\method}{\textbf{CRAFT }}
\section{Introduction}
\label{sec:introduction}
\begin{figure}[!t]
  \centering
  \includegraphics[width=\linewidth]{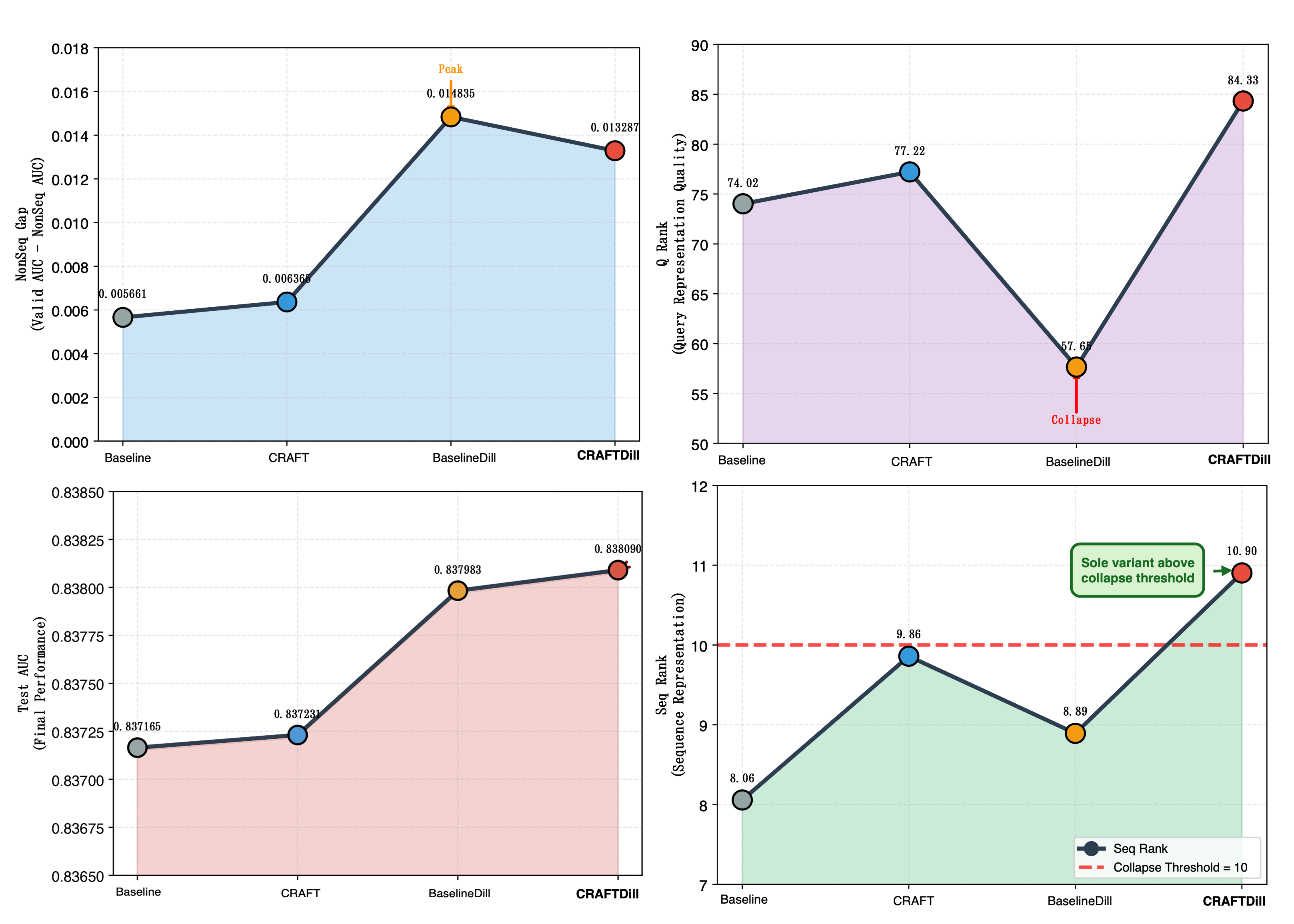}
  \caption{Effect of CRAFT and distillation on predictive performance and representation quality.}
  \label{fig:layerwise_seq_rank}
\end{figure}

Industrial CTR/CVR ranking models rely on both non-sequential multi-field features and sequential user behaviors. The former describe user profiles, item attributes, contextual signals, and cross features, and are commonly handled by feature-interaction networks \cite{guo2017deepfm,wang2017dcn,lian2018xdeepfm,zhu2025rankmixer}. The latter capture the temporal evolution of user interests and are usually modeled by target-aware or self-attentive sequence encoders \cite{zhou2018din,zhou2019dien,kang2018sasrec,sun2019bert4rec,qi2020sim,chai2025longer}. However, many industrial systems still connect these two sources through separated modules or late-stage fusion, which limits intermediate feature-sequence exchange and weakens scaling benefits.Recent unified recommendation architectures address this issue by placing sequential and non-sequential signals in a shared modeling framework. InterFormer, OneTrans, and HyFormer strengthen feature-sequence interaction through unified blocks \cite{zhang2026onetrans,huang2026hyformer,zeng2025interformer}. HyFormer is especially relevant to query-based unified modeling: it generates global query tokens from non-sequential features, decodes long-sequence states with these queries, and updates the queries through token mixing. These works suggest that unified blocks are a natural path for scaling recommendation models \cite{zhang2024wukong,zhu2025rankmixer,chai2025longer}.

\begin{figure*}[!t]
  \centering
  \includegraphics[width=\linewidth]{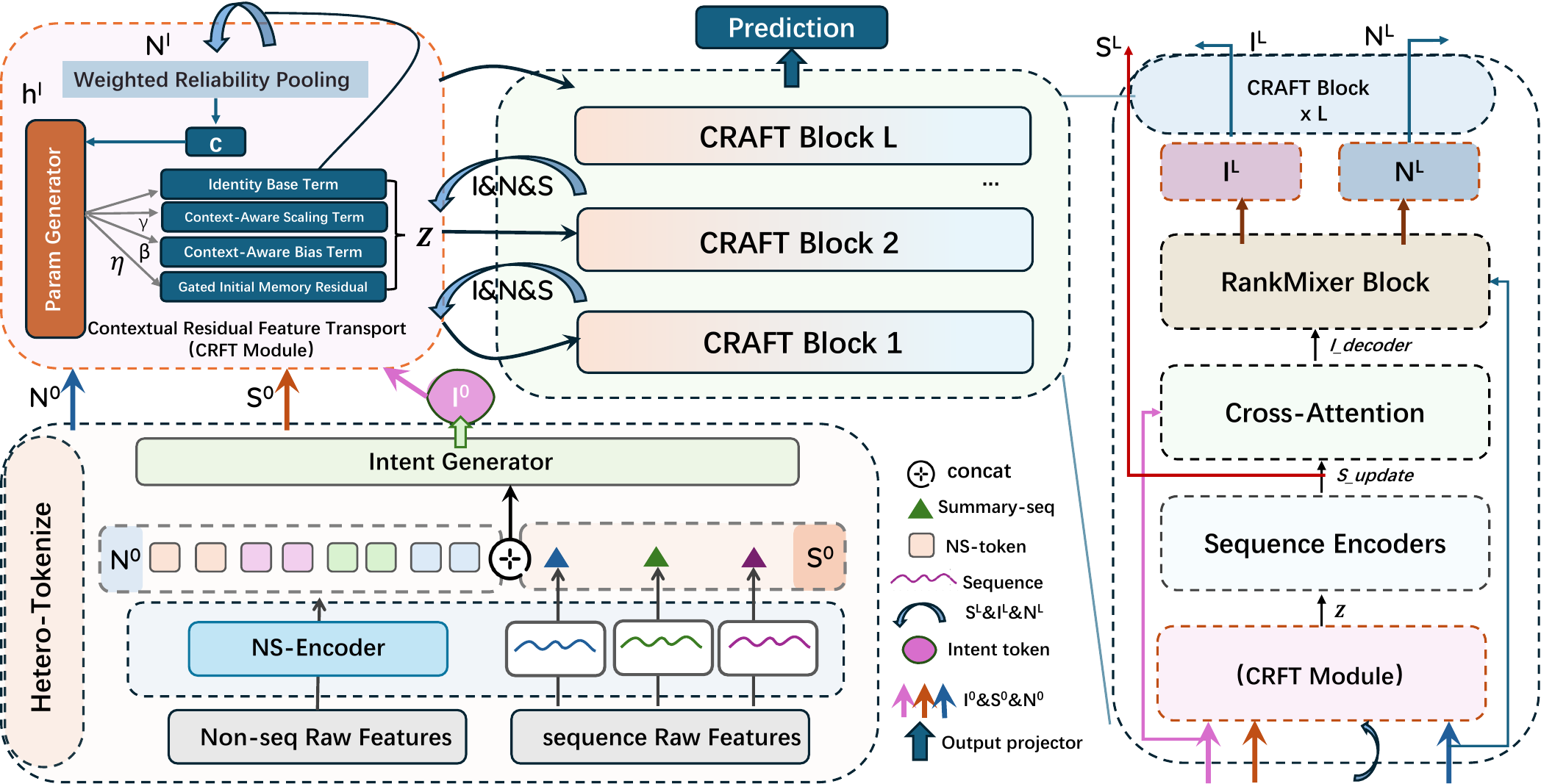}
  \caption{Overall architecture of CRAFT. Heterogeneous non-sequential and sequential features are first tokenized into non-sequential tokens, sequence summaries, and intent tokens. Each stacked CRAFT block performs contextual residual feature transport, where non-sequential context generates adaptive residual transformations to guide intent and sequence representations before unified interaction. This design enables layer-wise feature transport across heterogeneous signals while preserving stackability for depth scaling.}
  \label{fig:illustrate}
\end{figure*}

Nevertheless, interaction-based unification alone is not enough. Recommendation inputs are highly heterogeneous: dense vectors, sparse categorical fields, contextual signals, and behavior tokens differ in variance, frequency, reliability, and semantic granularity. Some fields may induce collapsed embedding spaces in large recommendation models \cite{guo2023embeddingcollapse}. In unified architectures, such instability can propagate through shared operators. TokenFormer reports a related phenomenon where collapsed non-sequential fields degrade sequential representations \cite{zhou2026tokenformer}. In query-based models, the bottleneck appears at the intent-carrier level: unreliable non-sequential representations may dominate query updates, dilute sequence evidence, and reduce intent diversity.

We argue that scalable unified recommendation should move from feature interaction to \emph{feature transport}. Instead of only asking how heterogeneous tokens interact within a layer, we ask how contextual information is carried, filtered, and preserved across stacked blocks. Inspired by flow-based representation dynamics, we view a deep unified model as a discrete context-conditioned representation evolution process. Each block should predict residual displacement for intent and sequence states while preserving useful memory from earlier layers, without requiring ODE solvers, stochastic paths, or flow-matching objectives.To this end, we propose \method, a unified and stackable block for scalable CTR/CVR prediction. \method summarizes non-sequential fields into a reliability-aware contextual field, which generates adaptive residual displacement and memory-preserving gates for intent and sequence representations before bridge-based information exchange. Thus, non-sequential context becomes an active controller of representation evolution rather than a passive object of interaction.
Our contributions are summarized as follows:
\begin{itemize}
    \item We identify a feature transport bottleneck in unified recommendation: even when sequential and non-sequential features can interact, useful intent information may still be overwritten, compressed, or destabilized during layer-wise propagation. This provides a new explanation for why simply increasing interaction depth or model size does not always improve ranking performance.

    \item We propose \method, a unified feature transport block that changes the role of non-sequential features from passive interaction targets to active controllers of representation evolution. By generating context-conditioned residual displacement and memory-preserving gates, \method explicitly regulates how non-sequential context guides intent and sequence representations across depth.

    \item We instantiate this idea with reliability-aware context aggregation and stackable contextual residual transport, enabling heterogeneous sparse, dense, and sequential signals to be jointly modeled without collapsing them into an uncontrolled homogeneous token space.

    \item We conduct extensive competition-scale experiments on TAAC2026. \method achieves a test AUC of 0.838090, surpassing the previous leaderboard-best score of 0.83798, and further reaches 0.838148 when scaled to six blocks. Representation diagnostics show that CRAFT improves scaling by selectively expanding the intent-carrier transport space rather than merely increasing final-layer capacity.
\end{itemize}
\section{Related Work}

\subsection{Feature Interaction and Sequential Recommendation}

Industrial CTR/CVR prediction has mainly developed along two lines. Feature interaction models capture sparse cross-field signals, from factorization machines and field-aware variants \cite{rendle2010fm,juan2016ffm}, to neural CTR models with memorization, product interactions, and FM-based layers \cite{cheng2016wide,guo2017deepfm,qu2016pnn,he2017nfm}. Later architectures model higher-order crosses with compressed interaction networks, explicit cross layers, bilinear weighting, or self-attentive field interaction \cite{lian2018xdeepfm,wang2017dcn,wang2021dcnv2,huang2019fibinet}.

Sequential recommendation instead encodes user behavior histories. Early methods use recurrent or convolutional encoders \cite{hidasi2016gru4rec,tang2018caser}. Target-aware CTR models retrieve candidate-relevant behaviors \cite{zhou2018din,zhou2019dien}, while Transformer-based methods capture longer-range dependencies \cite{chen2019bst,kang2018sasrec,sun2019bert4rec}. Recent industrial systems further extend history length through lifelong retrieval, real-time action modeling, and ultra-long sequence scaling \cite{chang2023twin,xia2023transact,si2024twinv2,chai2025longer}. However, these two lines are often connected by late fusion, limiting intermediate information flow between multi-field context and behavior sequences.

\subsection{Unified Recommendation Backbones}

Recent work moves toward unified backbones that jointly model sequential and non-sequential signals. Interaction-centric models suggest the benefit of joint field modeling \cite{wang2021dcnv2,zhou2018din,song2019autoint}. HiFormer and InterFormer use Transformer-style operators or explicit interaction tokens for heterogeneous feature interaction \cite{gui2023hiformer,zeng2025interformer}. OneTrans unifies feature interaction and sequence modeling in one Transformer framework \cite{zhang2026onetrans}. HyFormer introduces a query-based hybrid block for long-sequence and non-sequential feature modeling \cite{huang2026hyformer}, while TokenFormer studies shared token-stream unification and its robustness risks \cite{zhou2026tokenformer}. These works show the promise of stackable unified blocks, but most designs still rely on direct token mixing, attention, or query decoding as the main mechanism for connecting heterogeneous features.

\subsection{Scaling and Representation Collapse}

Scaling recommendation models requires balancing capacity, computation, and representation quality. Industrial ranking systems such as LiRank study production-scale modeling constraints \cite{borisyuk2024lirank}, while Wukong and Rank-Mixer explore scalable and efficient ranking backbones \cite{zhang2024wukong,zhu2025rankmixer}. Other studies analyze scaling in sequential recommendation or efficient large recommendation models \cite{zivic2024scalingseq,khrylchenko2025scaling,xu2025climber,kang2018sasrec}. Yet larger capacity does not necessarily yield richer representations: recommendation embeddings may collapse into low-dimensional subspaces \cite{guo2023embeddingcollapse}, and advertising recommendation can suffer from collapsed or entangled representations under heterogeneous sparse signals \cite{pan2024collapsedentangled}. TokenFormer further connects this risk to unified token-stream modeling \cite{zhou2026tokenformer}, suggesting that robust scaling needs mechanisms beyond simply mixing more heterogeneous tokens.

\section{Problem Setup and Preliminaries}

We study post-click conversion rate prediction in large-scale advertising recommendation. Each instance contains heterogeneous information from multiple sources, including sparse user and item fields, dense vector features, contextual signals, and multi-domain behavior sequences. Formally, an instance is denoted as
\begin{equation}
x =
\left[
\underbrace{
\mathcal{F}_u,\mathcal{F}_i,
\mathcal{D}_u,\mathcal{D}_i
}_{\text{non-sequential features}},
\underbrace{
\mathcal{S}_1,\ldots,\mathcal{S}_M
}_{\text{multi-domain behavior sequences}}
\right].
\end{equation}
where $\mathcal{F}_u,\mathcal{F}_i$ denote sparse user and item fields, $\mathcal{D}_u,\mathcal{D}_i$ denote dense user and item features, and $\mathcal{S}_m$ is the $m$-th behavior sequence. The goal is to learn a scoring function $f_\theta(x)\in[0,1]$ for ranking candidate items. The model is trained with binary cross-entropy and evaluated by AUC.

The core challenge is to jointly model sequential and non-sequential signals. Behavior sequences describe dynamic user interests and require temporal or order-aware modeling. Non-sequential multi-field features provide global user/item priors, contextual cues, and dense semantic signals. These two sources are complementary but statistically heterogeneous: sequence tokens are high-variance and order-dependent, while many non-sequential fields are sparse, low-frequency, or distributionally unstable.
A common unified modeling strategy is to map all features into tokens and perform direct token interaction through attention or MLP layers. Although this avoids late fusion, it also introduces a new difficulty: heterogeneous tokens are repeatedly mixed by shared operators. As the model becomes deeper or wider, unstable non-sequential fields may dominate query updates, dilute sequence evidence, or cause representation collapse. Thus, scalable unified recommendation requires not only stronger interaction, but also a controlled mechanism for carrying contextual information across stacked blocks.

This motivates our design goal: a unified block should preserve the benefits of token-based modeling while controlling how non-sequential context influences sequence-aware representations over depth. The next section introduces CRAFT, which realizes this goal through contextual residual adaptive feature transport.
\section{Method}

We propose CRAFT, short for Contextual Residual Adaptive Feature Transport,
a unified and stackable block for scalable recommendation. CRAFT is motivated
by a shift from feature interaction to feature transport. Feature interaction
focuses on how heterogeneous features are mixed inside each layer, for example
through attention, MLPs, or cross networks. This view treats non-sequential
fields mainly as objects to be fused with sequence representations. However,
in deep unified recommendation models, the key issue is not only whether two
features can interact, but also how contextual information is carried, filtered,
and preserved across stacked blocks. We call this process feature transport.

In CRAFT, non-sequential multi-field features are summarized into a contextual
field. This field does not simply participate in one-shot fusion. Instead, it
generates layer-wise residual displacement and memory-preserving signals that
guide the evolution of intent and sequence representations. Therefore,
non-sequential context changes from a passive object of interaction to an
active controller of representation evolution.

Given an input instance $x$, CRAFT first maps heterogeneous raw features into
three groups of tokens:
\begin{equation}
\mathcal{X}^{(0)}
=
\left(
\underbrace{\{\mathbf{I}^{(0)}_m\}_{m=1}^{M}}_{\text{intent tokens}},
\underbrace{\mathbf{N}^{(0)}}_{\text{non-sequential tokens}},
\underbrace{\{\mathbf{S}^{(0)}_m\}_{m=1}^{M}}_{\text{sequence tokens}}
\right).
\end{equation}
At layer $l$, CRAFT maintains the same three groups:
\begin{equation}
\mathcal{X}^{(l)}
=
\left(
\{\mathbf{I}^{(l)}_m\}_{m=1}^{M},
\mathbf{N}^{(l)},
\{\mathbf{S}^{(l)}_m\}_{m=1}^{M}
\right),
\end{equation}
where $\mathbf{I}^{(l)}_m$ are intent tokens for the $m$-th behavior domain,
$\mathbf{N}^{(l)}$ are non-sequential multi-field tokens, and
$\mathbf{S}^{(l)}_m$ are sequence tokens. Intent tokens act as compact
instance-specific states that accumulate, filter, and transport information
between global multi-field context and dynamic behavior evidence.

Each CRAFT block has two stages. First, contextual residual transport uses
non-sequential tokens to modulate intent and sequence states. Second,
bridge-based exchange updates the transported states through sequence
encoding, intent-to-sequence retrieval, and token mixing:
\begin{equation}
\widetilde{\mathcal{X}}^{(l)}
=
\mathcal{T}^{(l)}(\mathcal{X}^{(l)}),
\qquad
\mathcal{X}^{(l+1)}
=
\mathcal{B}^{(l)}(\widetilde{\mathcal{X}}^{(l)}).
\end{equation}
This decomposition makes CRAFT different from interaction-only designs:
non-sequential features are not merely tokens to be mixed, but contextual
controllers that shape the layer-wise movement of other representations.

\subsection{Heterogeneous Tokenization}

CRAFT first maps raw sparse, dense, paired, contextual, and sequential features
into a shared $d$-dimensional token space. Sparse user and item fields are
compressed into a fixed number of non-sequential tokens by RankMixer-style
tokenizers. Dense vectors are processed by structure-aware tokenizers: ordinary
vectors are projected by dense MLP tokenizers, segmented vectors are pooled
with masks, and structured dense features are decomposed according to their
internal layout. Paired sparse-dense features are encoded by pair tokenizers
that explicitly model ID-vector interactions.

The resulting non-sequential tokens are denoted as:
\begin{equation}
\mathbf{N}^{(0)}
=
[
\mathbf{N}_{u}^{sp},
\mathbf{N}_{u}^{de},
\mathbf{N}_{i}^{sp},
\mathbf{N}_{i}^{de},
\mathbf{n}_{time}
].
\end{equation}
Behavior sequences are embedded into $M$ domain-specific streams
$\{\mathbf{S}^{(0)}_m\}_{m=1}^{M}$. For each behavior domain, CRAFT constructs
intent tokens from both non-sequential context and the corresponding sequence
summary. Specifically, we first pool each sequence stream into a compact
summary $\bar{\mathbf{s}}_m$, concatenate it with the flattened non-sequential
tokens, and pass the result through a small intent generator:
\begin{equation}
\mathbf{I}^{(0)}_m
=
g_{\mathrm{int}}
\left(
\operatorname{Concat}
(
\operatorname{Flatten}(\mathbf{N}^{(0)}),
\bar{\mathbf{s}}_m
)
\right),
\quad m=1,\ldots,M .
\end{equation}
where $g_{\mathrm{int}}(\cdot)$ is an MLP-based generator that outputs
fixed-size intent tokens initialized from both global non-sequential context
and domain-specific behavior evidence.

\subsection{Contextual Residual Feature Transport}

The transport module is the core of CRAFT. At each layer, CRAFT summarizes
non-sequential tokens into a contextual field. Instead of simple average
pooling, we use reliability-weighted pooling so that different fields can
contribute unequally:
\begin{equation}
\mathbf{c}^{(l)}
=
\sum_{k=1}^{K}
a_k^{(l)}\mathbf{n}^{(l)}_k,
\qquad
a_k^{(l)}
=
\operatorname{softmax}_k
r^{(l)}(\mathbf{n}^{(l)}_k).
\end{equation}
Here $r^{(l)}(\cdot)$ is a small reliability scorer. The context vector
$\mathbf{c}^{(l)}$ then generates transport parameters for intent and sequence
streams:
\begin{equation}
(\boldsymbol{\gamma}^{(l)}_Z,\boldsymbol{\beta}^{(l)}_Z,\boldsymbol{\eta}^{(l)}_Z)
=
h_Z^{(l)}(\mathbf{c}^{(l)}),
\quad Z\in\{I,S\}.
\end{equation}

For a target representation $\mathbf{Z}^{(l)}_m$, where $\mathbf{Z}$ can be
either intent or sequence tokens, CRAFT applies:
\begin{equation}
\widetilde{\mathbf{Z}}^{(l)}_m
=
\mathbf{Z}^{(l)}_m
+
\boldsymbol{\gamma}^{(l)}_Z
\odot
\operatorname{RMSNorm}(\mathbf{Z}^{(l)}_m)
+
\boldsymbol{\beta}^{(l)}_Z
+
\boldsymbol{\eta}^{(l)}_Z
\odot
(\mathbf{Z}^{(0)}_m-\mathbf{Z}^{(l)}_m).
\end{equation}
The first residual term performs context-conditioned displacement, while the
last term preserves a gated path to the initial representation. This design
lets non-sequential context continuously guide intent and sequence states
without overwriting their original information. This operation can also be
viewed as a discrete context-conditioned transport step: each block predicts
a residual displacement for intent and sequence states conditioned on
non-sequential context. Unlike generative flow models, CRAFT does not use ODE
solvers or flow-matching losses; it keeps standard discriminative training
while making representation evolution more controllable across depth.

\subsection{Bridge-based Information Exchange}

After contextual transport, CRAFT performs explicit information exchange. The
bridge operator has three parts. First, each behavior sequence is refined by a
lightweight sequence encoder. Second, transported intent tokens retrieve
behavior evidence through intent-to-sequence cross-attention. Third, decoded
intent tokens and non-sequential tokens are jointly mixed by RankMixer, which
rewires token subspaces and applies normalized SwiGLU channel mixing.

We write the bridge compactly as:
\begin{equation}
(\mathbf{I}^{(l+1)},\mathbf{N}^{(l+1)},\mathbf{S}^{(l+1)})
=
\mathcal{B}^{(l)}
(
\widetilde{\mathbf{I}}^{(l)},
\mathbf{N}^{(l)},
\widetilde{\mathbf{S}}^{(l)}
).
\end{equation}
Unlike late-fusion systems, this bridge is applied at every layer. Therefore,
sequence evidence, intent states, and non-sequential context are repeatedly
aligned through stacked CRAFT blocks.

\subsection{Prediction and Training}

After $L$ CRAFT blocks, intent tokens from all behavior domains are
concatenated and projected into a final ranking representation:
\begin{equation}
\mathbf{z}
=
\operatorname{Proj}
\left(
[
\mathbf{I}^{(L)}_1,\ldots,\mathbf{I}^{(L)}_M
]
\right),
\qquad
\hat{y}
=
\sigma(\operatorname{MLP}(\mathbf{z})).
\end{equation}
The model is trained with binary cross-entropy. In the distilled setting, we
further match teacher soft predictions:
\begin{equation}
\mathcal{L}
=
(1-\alpha)\mathcal{L}_{\mathrm{BCE}}
+
\alpha\mathcal{L}_{\mathrm{KD}}.
\end{equation}
Distillation is used as a training enhancement and is kept separate from the
architectural contribution of CRAFT.

\begin{table*}[t]
\centering
\caption{Main results on TAAC2026. $\Delta$ denotes the AUC improvement over the original baseline, measured in per mille (\textperthousand). Params are counted at inference time; GFLOPs exclude embedding lookup.}
\label{tab:main_results}
\begin{tabular}{llcccccc}
\toprule
Paradigm & Model & Test AUC & $\Delta$ & Params & GFLOPs & Intent-Rank & Seq-Rank \\
\midrule
Original
& Original Baseline & 0.832688 & -- & 182.21M & -- & -- & -- \\
\midrule
\multirow{2}{*}{Interaction}
& Optimized Baseline & 0.837165 & +4.48 & 307.27M & 6.955 & 74.02 & 8.06 \\
& Optimized Baseline + Distill & 0.837983 & +5.30 & 334.87M & 9.889 & 57.68 & 8.89 \\
\midrule
\multirow{2}{*}{Transport}
& CRAFT & 0.837231 & +4.54 & 308.69M & 6.970 & 77.22 & 9.86 \\
& \cellcolor{gray!10} CRAFT + Distill
& \cellcolor{gray!10} \textbf{0.838090}
& \cellcolor{gray!10} \textbf{+5.40}
& \cellcolor{gray!10} 308.69M
& \cellcolor{gray!10} 6.970
& \cellcolor{gray!10} \textbf{84.33}
& \cellcolor{gray!10} \textbf{10.90} \\
\bottomrule
\end{tabular}
\end{table*}
\section{Experiments}
\label{sec:experiments}

We evaluate CRAFT on the TAAC2026 advertising recommendation task. The experiments focus on three questions: whether feature transport improves a strong unified interaction model, whether it is complementary to distillation, and whether the resulting block can scale along depth and width.

\subsection{Experimental Setup}

\textbf{Dataset and metric.}
TAAC2026 contains large-scale user-item advertising interaction records with sparse user/item fields, dense vector features, contextual signals, and four groups of behavior sequences. The task is conversion-related ranking, and the official metric is test AUC.

\textbf{Compared methods.}
The \emph{Original Baseline} is the initial official-style baseline. The \emph{Optimized Baseline} is a strong HyFormer-style unified model with structure-aware tokenization, RankMixer-style non-sequential tokenization, stable optimization, and EMA. \emph{Optimized Baseline + Distill} further applies teacher-student distillation. \emph{CRAFT} replaces interaction-only propagation with contextual residual adaptive feature transport, and \emph{CRAFT + Distill} applies the same distillation setting to CRAFT.

\textbf{Implementation and evaluation.}
Unless otherwise specified, CRAFT uses hidden dimension $420$, embedding dimension $64$, four attention heads, two intent tokens per behavior domain, SwiGLU sequence encoders, and RankMixer-style non-sequential tokenization. Distillation uses temperature $1.0$ and weight $0.65$. Params are counted at inference time, and GFLOPs are measured per sample under the same sequence-length setting, excluding embedding lookup. We also report effective rank of intent tokens and sequence summaries as representation diagnostics.

\subsection{Main Results}

Table~\ref{tab:main_results} shows that CRAFT consistently improves over the interaction-based baseline. Without distillation, CRAFT increases test AUC from $0.837165$ to $0.837231$, indicating that feature transport provides gains beyond tokenization and optimization. With distillation, CRAFT reaches $0.838090$, surpassing the distilled interaction baseline while using fewer inference-time parameters and lower measured GFLOPs. The higher Intent-Rank and Seq-Rank further suggest that feature transport preserves richer intermediate intent and sequence representations.

\subsection{Scaling Results}

\begin{table}[t]
  \centering
  \caption{Depth and width scaling of CRAFT. GFLOPs are measured per sample and exclude embedding lookup.}
  \label{tab:craft_scaling}
  \small
  \setlength{\tabcolsep}{4.5pt}
  \begin{tabular}{lcccc}
    \toprule
    Model & Blocks & $d_{\mathrm{model}}$ & Test AUC & GFLOPs \\
    \midrule
    CRAFT-1L & 1 & 420 & 0.837975 & 3.960 \\
    CRAFT-2L & 2 & 420 & 0.838090 & 6.970 \\
    CRAFT-4L & 4 & 420 & 0.838093 & 12.992 \\
    CRAFT-6L & 6 & 420 & \textbf{0.838148} & 19.013 \\
    \midrule
    CRAFT-D600 & 2 & 600 & 0.838106 & 13.679 \\
    \bottomrule
  \end{tabular}
\end{table}

Table~\ref{tab:craft_scaling} evaluates scaling along depth and width. Increasing CRAFT from one to six blocks improves test AUC from $0.837975$ to $0.838148$, showing that the transport block remains stackable under larger depth. Increasing the hidden dimension from $420$ to $600$ at two blocks also improves AUC from $0.838090$ to $0.838106$. These results indicate that CRAFT benefits from both vertical and horizontal scaling, with the best current result obtained by six stacked blocks.
\begin{figure}[!h]
  \centering
 \includegraphics[width=\linewidth]{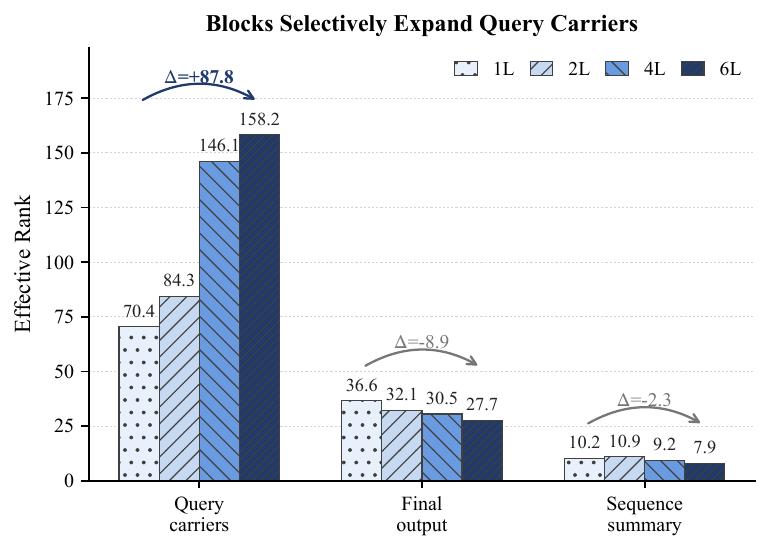}
    \caption{Block-stacking diagnostic for CRAFT. We compare the effective ranks of query carriers, final outputs, and sequence summaries when increasing the number of CRAFT blocks from 1 to 6. Query carriers show a large rank increase from 70.4 to 158.2, while the final-output rank decreases and the sequence-summary rank remains low. This indicates that depth scaling mainly expands the query-mediated transport space rather than uniformly increasing the dimensionality of all representation spaces.}
    \label{fig:block_scaling_rank}
\end{figure}
\subsection{Representation Diagnostics}

Representation diagnostics explain why CRAFT scales. Compared with interaction-only models, CRAFT improves the effective rank of intent and sequence representations, suggesting that contextual transport preserves more usable intermediate information. Additional visualizations and layer-wise analyses are provided in Appendix~\ref{app:representation_diagnostics}.

\section*{Ethics and Privacy Statement}

This work studies ranking models for large-scale advertising recommendation. While improved recommendation accuracy may benefit user experience and platform efficiency, such models may also amplify historical bias, over-personalization, or uneven exposure across user and item groups. We use anonymized competition data and do not attempt to identify individuals; future deployment should include privacy-preserving data governance, fairness auditing, and monitoring for distribution shift and misuse.
\section*{Acknowledgements}
This work was supported by the National Science and Technology Major Project for New Generation Artificial Intelligence under Grant No. 2025ZD0122702, and by the Tencent Rhino-Bird WeChat Special Project under Grant No. WXG-FR-2026-11. We also thank Tencent for organizing the TAAC competition and the Tencent UNI-REC Challenge.



\bibliographystyle{ACM-Reference-Format}
\bibliography{ref}

\appendix
\begin{table*}[!htbp]
\centering
\caption{Feature groups in the TAAC2026 advertising recommendation data.}
\label{tab:dataset_features}
\begin{tabular}{llll}
\toprule
Group & Raw form & Count & Model-side token type \\
\midrule
Basic columns & IDs and timestamps & 5 & Supervision/context fields \\
User integer fields & Scalar IDs / ID arrays & 54 & User non-sequential tokens \\
User dense fields & Float vectors / arrays & 17 & Dense and pair-aware tokens \\
Item integer fields & Scalar IDs / ID arrays & 17 & Item non-sequential tokens \\
Item dense fields & Float vectors & 4 & Item dense tokens \\
Behavior sequences & Ordered ID sequences & 45 in 4 domains & Sequence tokens \\
\bottomrule
\end{tabular}
\end{table*}
\section{Dataset Details and Feature Processing}
\label{app:data_features}
The TAAC2026 advertising recommendation data is organized as large-scale user-item interaction records. According to the official public schema description\footnote{\url{https://huggingface.co/datasets/TAAC2026/second_round_sample_1000}}, each instance contains five basic columns: user ID, item ID, label type, label time, and request timestamp. The remaining inputs are heterogeneous recommendation signals, including user-side integer fields, user-side dense fields, item-side integer fields, item-side dense fields, and multi-domain behavior sequences. Table~\ref{tab:dataset_features} summarizes the major feature groups.

Table~\ref{tab:dataset_features} shows that the task input is not a homogeneous token stream. Sparse integer fields, dense numerical vectors, structured dense arrays, paired ID-dense arrays, and ordered behavior sequences have different statistical properties. Sparse fields are high-cardinality and unevenly updated. Dense fields may contain continuous embeddings, repeated segments, metadata-like suffixes, or missing-value conventions. Behavior sequences are order-dependent and should preserve temporal structure. Therefore, a unified model cannot simply concatenate all raw fields and rely on a downstream block to discover every structure automatically.

This observation motivates the first design layer of our system: heterogeneous tokenization. Before CRAFT performs feature transport, different raw feature families are converted into stable token representations. The goal is not to manually inject semantic labels into anonymized fields, but to preserve their observed structural patterns and avoid creating noisy tokens that harm scaling.

\begin{table}[!htbp]
\centering
\caption{Tokenization pipeline for heterogeneous recommendation features.}
\label{tab:appendix_tokenization}
\setlength{\tabcolsep}{1.5pt}
\begin{tabular}{llll}
\toprule
Group & Raw form & Encoder & Output \\
\midrule
User IDs & Sparse IDs & RankMixer & User tokens \\
Item IDs & Sparse IDs & RankMixer & Item tokens \\
Multi-value IDs & ID arrays & Masked pool & Field tokens \\
Dense fields & Vectors & MLP / structured & Dense tokens \\
ID-dense pairs & Paired arrays & Pair encoder & Pair tokens \\
Behaviors & Sequences & Seq encoder & Seq tokens \\
Context & Time fields & Embedding & Context token \\
\bottomrule
\end{tabular}
\end{table}

Table~\ref{tab:appendix_tokenization} summarizes the tokenization pipeline. User and item integer fields are embedded and compressed into non-sequential tokens. Multi-value ID arrays are aggregated with masks so that padded positions do not become artificial evidence. Dense vectors are processed by either simple MLP projection or structure-aware tokenizers depending on their observed layout. Behavior histories remain sequence tokens because their order carries essential information for intent modeling. Timestamp-derived signals are encoded as contextual tokens.

This tokenization stage is important for the feature transport view. CRAFT assumes that non-sequential tokens can provide a contextual field that guides intent and sequence states. If the contextual tokens themselves are noisy or structurally misaligned, the transport path can amplify bad evidence. In contrast, once heterogeneous fields are converted into reliable tokens, CRAFT can use non-sequential context as a stable controller of representation evolution.

\subsection{Structure-Aware Dense Tokenization}
\label{app:dense_tokenization}

\begin{table*}[!htbp]
\centering
\caption{Structure-aware dense tokenization. Dense fields are grouped by observed structural pattern rather than anonymized field ID.}
\label{tab:appendix_dense_tokenizers}
\begin{tabular}{llll}
\toprule
Pattern & Observed structure & Encoder & Output \\
\midrule
Plain vector & Continuous vector & BN + MLP & 1 token \\
Segmented vector & Repeated sub-vectors & Masked segment pool & 1 token \\
Positioned segments & Position + segments + tail & Pos + segment + tail fusion & 1 token \\
Embedding + metadata & Vector prefix + suffix & Prefix projection & 1 token \\
Item dense & Item-side vectors & Field pool & 1 token \\
ID-dense pair & Aligned ID/value arrays & Pair encoder & Field token \\
\bottomrule
\end{tabular}
\end{table*}

Table~\ref{tab:appendix_dense_tokenizers} reports the dense-tokenization rules used in our system. The key observation is that anonymized dense vectors are not guaranteed to be ordinary continuous vectors. Some dense features behave like stable embedding vectors; some contain repeated fixed-size sub-vectors; some contain all-zero padded segments; some combine vector blocks with position-like metadata or low-dimensional tail statistics; and some are aligned with ID arrays. Treating all of them as flat vectors can mix semantically different subspaces and make the resulting token unstable.

For plain dense vectors, batch normalization followed by an MLP projection is sufficient because the vector behaves as one continuous feature. For segmented dense vectors, all-zero segments often indicate padding or missing entries. We therefore project each valid segment and aggregate them with a mask. This prevents zero segments from distorting normalization and pooling. For positioned segmented vectors, the position code, segment body, and tail statistics play different roles; we encode them separately and fuse them into one token. For stable embedding vectors with metadata suffixes, we use the stable vector prefix when the suffix appears distribution-sensitive. This avoids letting a small unstable suffix dominate a high-dimensional useful embedding.

For item dense fields, per-field projection followed by pooling is used because item dense features are fewer but semantically distinct. For ID-dense pairs, the dense value at each position is meaningful only together with the ID at the same position. The pair encoder therefore models ID embeddings, dense projections, multiplicative interactions, and difference features before masked aggregation.

This design supports the central claim of the paper. CRAFT is not merely a larger interaction layer; it depends on reliable feature transport. Reliable transport requires that dense fields first be converted into tokens whose internal structure is respected. Otherwise, increasing depth or width may simply transport dense-feature noise more strongly.

\subsection{Missing Values, Padding, and Sparse Robustness}
\label{app:missingness}
\begin{table}[!htbp]
\centering
\caption{Robust handling of missingness and padding.}
\setlength{\tabcolsep}{1.5pt}
\label{tab:appendix_missing}
\begin{tabular}{lll}
\toprule
Feature type & Dataset pattern & Operation \\
\midrule
Scalar ID & Missing code & Reserved embedding \\
ID array & Padding code & Masked aggregation \\
Dense vector & Invalid values & Sanitization + norm \\
Segmented dense & Empty segment & Segment mask \\
Behavior sequence & Padded events & Sequence mask \\
ID-dense pair & Invalid pair & Pair-validity mask \\
Large-vocab ID & Oversized vocabulary & Hashing / skip rule \\
\bottomrule
\end{tabular}
\end{table}

Table~\ref{tab:appendix_missing} summarizes the missing-value and padding rules. Missingness is common in industrial recommendation logs. It can be a valid signal, but it can also create fake interactions if handled incorrectly. Scalar ID fields use reserved missing embeddings so that missingness can be learned as a categorical pattern. Multi-value ID arrays and behavior sequences use masks to ensure that padding positions do not enter pooling or attention as real events.

Dense features require additional care because NaN, infinite values, and all-zero segments can destabilize normalization layers. We sanitize invalid numerical values, normalize dense vectors, and apply explicit masks for segmented dense fields. For aligned ID-dense arrays, pair validity is determined at the aligned-position level. This allows the model to retain useful dense evidence when present while preventing padded pairs from contributing to the token.

These robustness rules are closely related to representation collapse. If missing values are treated as normal feature values, the model may learn spurious low-diversity directions shared by many samples. Such directions can dominate non-sequential context and then propagate through unified blocks. By explicitly modeling missingness and padding, the input token space becomes less prone to such collapse-prone artifacts.

\section{System Components and Training Protocol}
\label{app:training_protocol}

This section provides the system-level details behind the main results. Unless otherwise specified, controlled comparisons keep the data split, sequence lengths, tokenizer configuration, optimizer setting, EMA strategy, distillation teacher, distillation weight, and effective global batch fixed. Runs that require a different data protocol or effective batch due to memory constraints are treated as engineering explorations rather than clean scaling points.

\subsection{Optimization and Stabilization}

\begin{table*}[!htbp]
\centering
\caption{Optimization and stabilization settings.}
\label{tab:appendix_optimizer}
\begin{tabular}{lll}
\toprule
Component & Setting & Applied to \\
\midrule
Sparse optimizer & Adagrad & Sparse embeddings \\
Dense optimizer & MuonPlus/AdamW-style update & Dense parameters \\
EMA & Dense-parameter EMA & Validation/checkpoint selection \\
Embedding refresh & Early sparse reinitialization & Sparse embeddings \\
LR schedule & Step-wise cosine decay & Dense optimizer \\
Model selection & Validation-based early stopping & Whole model \\
\bottomrule
\end{tabular}
\end{table*}
\begin{figure}[!h]
  \centering
  \includegraphics[width=\linewidth]{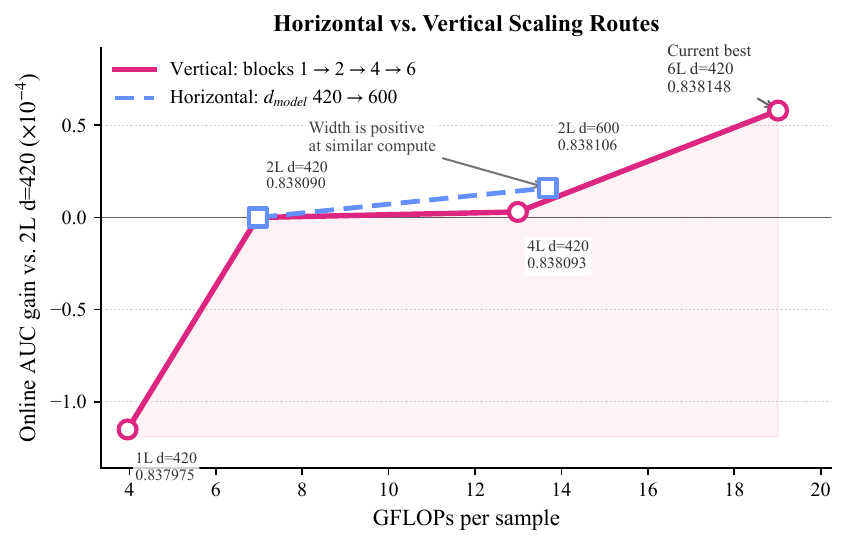}
  \caption{Comparison of horizontal and vertical scaling routes. Horizontal scaling increases $d_{\mathrm{model}}$ at fixed depth, while vertical scaling increases the number of CRAFT blocks at fixed $d_{\mathrm{model}}=420$.}
  \label{fig:scaling_route_comparison}
\end{figure}
Table~\ref{tab:appendix_optimizer} separates sparse and dense optimization. Sparse ID embeddings receive highly uneven and frequency-dependent gradients. Frequent IDs may receive stable updates, while rare IDs may receive only a few noisy gradients. Adagrad is therefore suitable because it adapts learning rates by coordinate frequency. Dense matrices, token mixers, sequence encoders, and CRAFT transport modules have different optimization geometry and are optimized with dense-parameter optimizers.

EMA is used to smooth the dense ranking function used for validation and checkpoint selection. This is especially useful in recommendation because validation AUC can fluctuate due to noisy labels and distribution shift. Early sparse reinitialization refreshes unreliable sparse embeddings at the beginning of training, reducing the chance that early noisy embedding states dominate later transport.

The broader lesson is that scaling a unified recommendation model is not only an architectural problem. Sparse memorization, dense representation learning, and sequence modeling have different training dynamics. A stable training protocol is necessary before evaluating whether a new block such as CRAFT can truly scale.

\subsection{Progressive Distillation}
\label{app:distillation}

In the distilled setting, the student is trained with both binary labels and teacher soft predictions:
\begin{equation}
\mathcal{L}
=
(1-\alpha)\mathcal{L}_{\mathrm{BCE}}
+
\alpha\mathcal{L}_{\mathrm{KD}},
\end{equation}
where $\mathcal{L}_{\mathrm{KD}}$ matches teacher and student probabilities with temperature $1.0$.

\begin{table*}[!htbp]
\centering
\caption{Progressive distillation protocol.}
\label{tab:appendix_distill}
\begin{tabular}{llll}
\toprule
Stage & Teacher source & Student initialization & Distillation weight \\
\midrule
Supervised anchor & None & Random & 0 \\
First-generation distillation & Strong supervised anchor & Random & 0.50 \\
Second-generation distillation & First-generation distilled model & Random & 0.65 \\
\bottomrule
\end{tabular}
\end{table*}

Table~\ref{tab:appendix_distill} describes the distillation protocol. Distillation is treated as a training enhancement rather than part of the CRAFT architecture. This distinction is important for interpretation. In noisy implicit-feedback recommendation, the binary label only gives partial preference information. Teacher probabilities provide smoother signals over difficult samples and encode relative uncertainty that is not visible in hard labels.

This explains why distillation improves both the interaction baseline and CRAFT. However, distillation and CRAFT affect the model differently. Distillation mainly improves the supervision signal, while CRAFT changes the internal representation pathway by controlling how non-sequential context is transported into intent and sequence states. The ablation in Appendix~\ref{app:ablations} separates these two effects.

\section{Ablation Studies}
\label{app:ablations}

\begin{table}[!htbp]
\centering
\caption{CRAFT ablation results. ``Trans'' indicates contextual feature transport; ``T-AUC'' denotes official test AUC.}
\label{tab:appendix_craft_ablation}
\begin{tabular}{lcccc}
\toprule
Model & Trans & Distill & T-AUC & Params \\
\midrule
Optim. Base & No & No & 0.837165 & 307.27M \\
CRAFT & Yes & No & 0.837231 & 308.69M \\
Optim. Base + Distill & No & Yes & 0.837983 & 334.87M \\
CRAFT + Distill & Yes & Yes & 0.838090 & 308.69M \\
\bottomrule
\end{tabular}
\end{table}

Table~\ref{tab:appendix_craft_ablation} isolates two sources of improvement: contextual feature transport and distillation. Without distillation, CRAFT improves the optimized interaction baseline from $0.837165$ to $0.837231$. This verifies that feature transport brings architectural value even when the training objective remains unchanged. The gain is small in absolute AUC but meaningful in this competition-scale setting, where the optimized baseline already includes strong tokenization, stable optimization, and unified feature-sequence interaction.

Distillation gives a larger improvement for both paradigms. The interaction baseline improves from $0.837165$ to $0.837983$, while CRAFT with the same distillation setting reaches $0.838090$. This shows that CRAFT is not redundant with distillation. If CRAFT merely duplicated the effect of a stronger teacher, its advantage would disappear after distillation. Instead, CRAFT still improves over the distilled interaction baseline.

The parameter comparison further strengthens this interpretation. CRAFT + Distill achieves the strongest main result with $308.69$M inference-time parameters, while the distilled interaction baseline uses $334.87$M parameters in our implementation. Therefore, the improvement is not explained by a larger inference model. It is better explained by how CRAFT changes the information path: non-sequential context becomes a controller of intent and sequence evolution, instead of only another object in an interaction layer.

\section{Representation Diagnostics}
\label{app:representation_diagnostics}

\begin{table*}[!htbp]
\centering
\caption{Representation diagnostics on sampled validation batches. Rank denotes effective rank; Cos denotes average absolute cosine similarity.}
\label{tab:appendix_representation}
\begin{tabular}{lccccccccc}
\toprule
Model & Test AUC & I-Rank & I-Cos & NS-Rank & NS-Cos & Seq-Rank & Seq-Cos & Final-Rank & Final-Cos \\
\midrule
Optimized Baseline & 0.837165 & 74.02 & 0.171 & 187.68 & 0.188 & 8.06 & 0.476 & 35.67 & 0.263 \\
CRAFT & 0.837231 & 77.22 & 0.210 & 180.98 & 0.127 & 9.86 & 0.485 & 33.29 & 0.263 \\
Optimized Baseline + Distill & 0.837983 & 57.68 & 0.225 & 207.44 & 0.210 & 8.89 & 0.477 & 47.55 & 0.239 \\
CRAFT + Distill & 0.838090 & 84.33 & 0.233 & 202.70 & 0.125 & 10.90 & 0.514 & 32.15 & 0.270 \\
\bottomrule
\end{tabular}
\end{table*}

Table~\ref{tab:appendix_representation} reports effective rank and cosine diagnostics for major representation spaces. The optimized baseline already has strong performance, but its intent and sequence spaces remain relatively limited. Adding CRAFT increases I-Rank from $74.02$ to $77.22$ and Seq-Rank from $8.06$ to $9.86$. This indicates that feature transport preserves more diverse intent directions and richer sequence evidence.

The non-sequential space behaves differently. CRAFT decreases NS-Cos from $0.188$ to $0.127$. This suggests that CRAFT does not simply make non-sequential features more dominant. Instead, it makes non-sequential context less redundant and uses it to modulate other states. This is consistent with the transport view: non-sequential features act as contextual controllers, while intent and sequence states become the main evolving representations.

Distillation changes the geometry in a different way. The distilled interaction baseline obtains strong AUC but has lower I-Rank than the non-distilled optimized baseline. This means that soft labels can improve final ranking quality without necessarily expanding the intermediate intent space. By contrast, CRAFT + Distill obtains both the best AUC and the highest I-Rank and Seq-Rank among the main models. This suggests a complementary mechanism: distillation improves supervision smoothness, while CRAFT improves representation transport.

Final-Rank should not be interpreted as ``the higher the better.'' The final representation is a readout space for producing a calibrated ranking score. It can be compact as long as the intermediate transport spaces preserve useful evidence. The more relevant signal for CRAFT is that intent and sequence representations become richer before final projection.
\subsection{KMeans-MI Analysis}
\label{app:kmeans_mi}
\begin{figure}[!htbp]
  \centering
  \includegraphics[width=\linewidth]{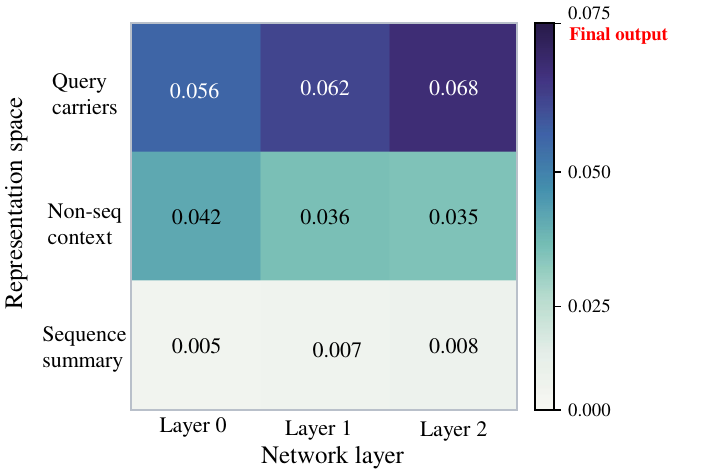}
  \caption{KMeans-MI heatmap of label-discriminative information across CRAFT representation spaces. KMeans clustering with $K=128$ is applied to each representation, and mutual information is computed between cluster assignments and labels.}
  \label{fig:kmeans_mi_heatmap}
\end{figure}

Figure~\ref{fig:kmeans_mi_heatmap} examines where label-discriminative information is encoded inside CRAFT. Intent carriers contain the strongest intermediate label signal, with MI increasing from $0.056$ at Layer 0 to $0.068$ at Layer 2. Sequence-summary representations remain much less informative, with MI below $0.009$ across layers. Non-sequential context representations provide moderate signal but gradually decrease from $0.042$ to $0.035$. The final output reaches the highest MI of $0.075$, as expected for the prediction space.

This result is important because it identifies the role of each representation group. Sequence summaries alone are not the primary label-discriminative space. Non-sequential context alone is also not the final answer. Instead, CRAFT progressively organizes label-relevant evidence through intent carriers. This supports the design of intent tokens as the transport interface between non-sequential context and sequential evidence.

The MI trend also helps distinguish CRAFT from simple feature interaction. In a pure interaction view, one may expect all useful information to be mixed directly into a final hidden vector. In CRAFT, useful information is first gathered and reorganized in intent carriers, then projected into the final prediction space. This explains why intermediate intent-rank expansion can coexist with a compact final readout.

\subsection{Qualitative Representation Geometry}
\label{app:qualitative_representation_geometry}

\begin{figure*}[!htbp]
    \centering
    \begin{subfigure}[t]{0.485\textwidth}
        \centering
        \includegraphics[width=\linewidth]{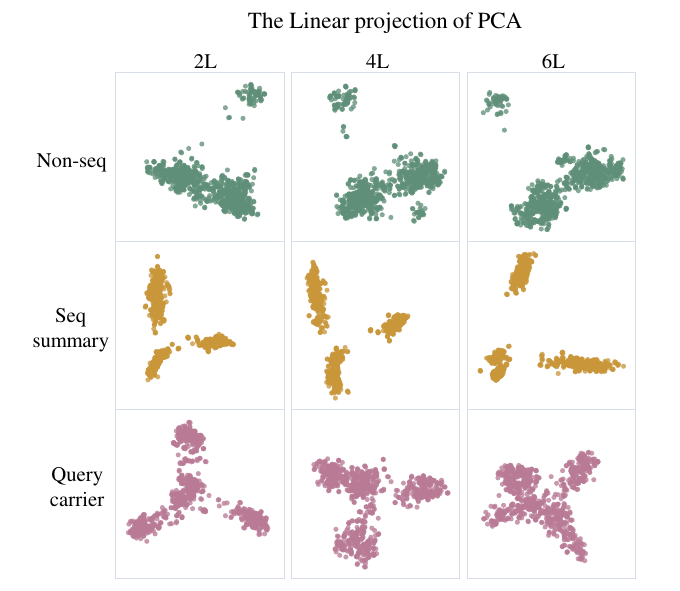}
        \caption{PCA projection.}
        \label{fig:supp_pca_representation}
    \end{subfigure}
    \hfill
    \begin{subfigure}[t]{0.46\textwidth}
        \centering
        \includegraphics[width=\linewidth]{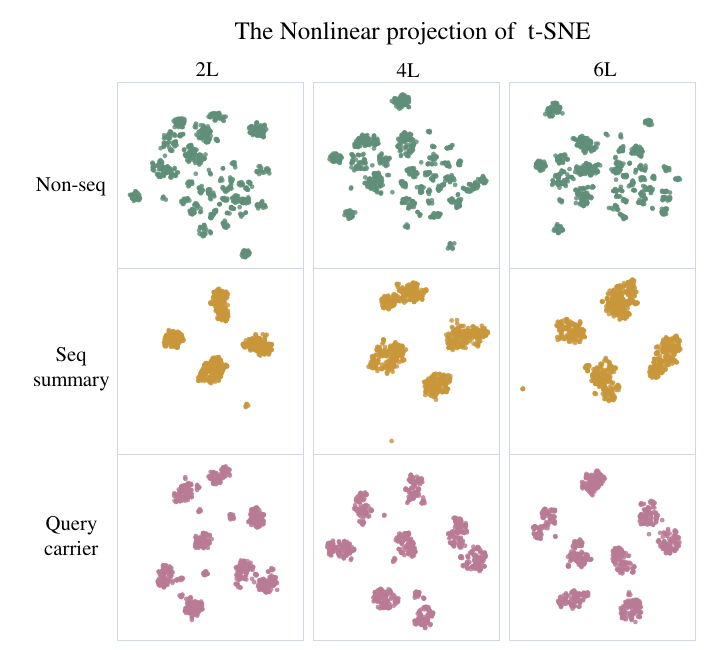}
        \caption{t-SNE projection.}
        \label{fig:supp_tsne_representation}
    \end{subfigure}
    \caption{Qualitative projections of layer-2 hidden representations in CRAFT-2L, CRAFT-4L, and CRAFT-6L. Rows correspond to non-sequential, sequence-summary, and intent-carrier representations, and columns correspond to model depth. Both PCA and t-SNE show that the clearest depth-dependent restructuring occurs in the intent-carrier space.}
    \label{fig:supp_representation_projection}
\end{figure*}

Figure~\ref{fig:supp_representation_projection} provides qualitative evidence for the representation changes observed in Table~\ref{tab:appendix_representation}. We visualize layer-2 states because this is the deepest shared post-transport state across CRAFT-2L, CRAFT-4L, and CRAFT-6L. The non-sequential representation space remains relatively stable across depths, while the sequence-summary space stays compact. The most visible geometric change occurs in the intent-carrier space.

These visualizations should not be interpreted as standalone proof of performance improvement. Their role is diagnostic. They show where the model changes when depth increases. The fact that the strongest geometric evolution occurs in intent carriers is consistent with the feature transport mechanism: additional CRAFT blocks mainly refine the interface that carries contextual and sequential evidence, instead of uniformly expanding every hidden representation.

\section{Scaling Analysis}
\label{app:scaling_analysis}

\begin{table}[!htbp]
\centering
\caption{Clean CRAFT scaling results.}
\label{tab:appendix_scaling}
\small
\setlength{\tabcolsep}{1.5pt}
\begin{tabular}{lccccc}
\toprule
Model & Blocks & $d_{\mathrm{model}}$ & Test AUC & Params & GFLOPs \\
\midrule
CRAFT-2L & 2 & 420 & 0.838090 & 308.69M & 6.970 \\
CRAFT-4L & 4 & 420 & 0.838093 & 330.63M & 12.992 \\
CRAFT-6L & 6 & 420 & 0.838148 & 352.56M & 19.013 \\
CRAFT-D600 & 2 & 600 & 0.838106 & 375.41M & 13.679 \\
\bottomrule
\end{tabular}
\end{table}

Table~\ref{tab:appendix_scaling} reports the clean scaling points. Along the depth axis, increasing CRAFT from one block to six blocks improves test AUC from $0.837975$ to $0.838148$. Along the width axis, increasing $d_{\mathrm{model}}$ from $420$ to $600$ at two blocks improves test AUC from $0.838090$ to $0.838106$. These results show that CRAFT can benefit from both vertical scaling and horizontal scaling.

The depth trend is positive but not perfectly linear. CRAFT-4L gives only a small improvement over CRAFT-2L, while CRAFT-6L gives a clearer gain. This suggests that stackability is not a trivial consequence of adding layers. Additional blocks must preserve useful intent information long enough for deeper refinement to help. The strong CRAFT-6L result supports the claim that contextual residual transport improves the stackability of unified recommendation blocks.

The width result is also useful. CRAFT-D600 improves over CRAFT-2L, showing that the transport block can absorb a wider hidden state. However, width scaling increases both parameter count and per-layer computation more aggressively than adding shallow depth. Therefore, the best route depends on the compute budget. In our current results, depth scaling to six blocks gives the best AUC, while width scaling provides complementary evidence that CRAFT uses additional capacity.

\subsection{Depth Scaling Expands Intent-Mediated Transport}
\label{app:depth_scaling_rank}

Figure~\ref{fig:block_scaling_rank} investigates why CRAFT benefits from stacking more blocks. When increasing the number of CRAFT blocks from 1 to 6, the effective rank of intent carriers increases from $70.4$ to $158.2$. In contrast, the final-output rank decreases from $36.6$ to $27.7$, and the sequence-summary rank remains low, changing from $10.2$ to $7.9$.

This contrast is central to the scaling interpretation. Deeper CRAFT does not uniformly inflate every representation space. Instead, additional blocks mainly expand the intent-mediated transport space. The final output can remain compact because it only needs to read out the transported evidence. The sequence-summary space can also remain compact because CRAFT does not rely on raw sequence summaries as the only information path. The model stores and refines cross-feature evidence in intent carriers.

This also explains why the feature transport view is more suitable than a pure feature interaction view for describing our model. The main effect of depth is not simply more pairwise mixing. It is the progressive expansion and refinement of the representation space that transports information between non-sequential context and behavior evidence.

\subsection{Width Scaling Increases Usable Representation Capacity}
\label{app:width_scaling_rank}

\begin{figure}[!htbp]
  \centering
  \includegraphics[width=\linewidth]{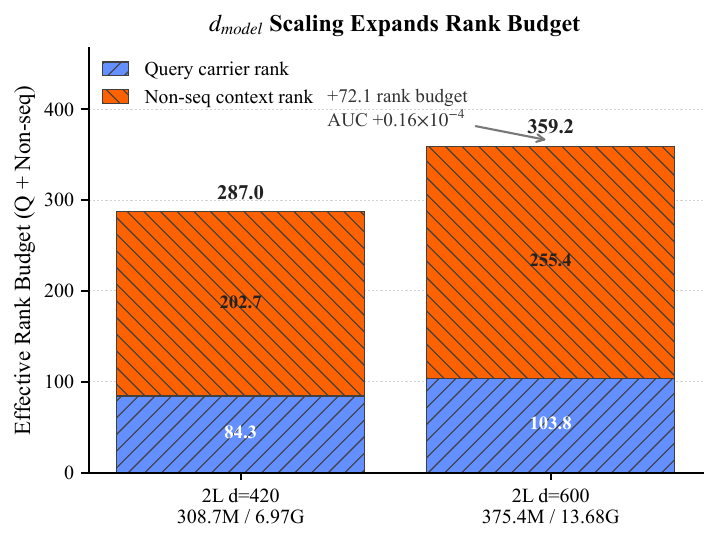}
  \caption{Width-scaling diagnostic for CRAFT. Increasing $d_{\mathrm{model}}$ from 420 to 600 increases the combined effective-rank budget of intent-carrier and non-sequential context representations.}
  \label{fig:dmodel_scaling_rank}
\end{figure}
\begin{table*}[!htbp]
\centering
\caption{Negative or limited-gain findings from controlled explorations.}
\label{tab:appendix_negative}
\begin{tabular}{lll}
\toprule
Direction & Controlled change & Observed outcome \\
\midrule
Dense tokenization & Flat projection for structured dense vectors & Weaker scaling behavior \\
Segmented dense fields & Remove segment-aware pooling & Lower robustness \\
Stable dense embeddings & Use metadata suffix with vector prefix & Worse generalization \\
Dense tokenizer complexity & Add heavier self-attention tokenizer & Higher cost and instability \\
Optimization & Use one optimizer for all parameters & Weaker stability \\
FFN expansion & Increase hidden multiplier & Limited or negative gain \\
Training duration & Continue after early best epoch & Overfitting trend \\
\bottomrule
\end{tabular}
\end{table*}

Figure~\ref{fig:dmodel_scaling_rank} studies whether increasing $d_{\mathrm{model}}$ provides usable capacity rather than merely adding parameters. Under the same 2-layer CRAFT architecture, increasing $d_{\mathrm{model}}$ from $420$ to $600$ raises intent-carrier rank from $84.3$ to $103.8$ and non-sequential context rank from $202.7$ to $255.4$. The combined rank budget increases from $287.0$ to $359.2$, together with a positive AUC change from $0.838090$ to $0.838106$.

This suggests that the added width is absorbed by meaningful intermediate representation spaces. The larger model does not only increase parameter count; it also expands the capacity of the context and intent spaces that participate in feature transport. This is useful evidence for width scaling, even though the absolute AUC gain is smaller than the best depth-scaling result.
\subsection{Horizontal and Vertical Scaling Routes}
\label{app:scaling_routes}
Figure~\ref{fig:scaling_route_comparison} compares horizontal and vertical scaling under the compute axis. CRAFT-1L is a strong low-compute starting point, but CRAFT-2L improves AUC from $0.837975$ to $0.838090$, showing that the second transport block is useful. Beyond CRAFT-2L, both scaling routes yield positive gains. At a similar compute level, CRAFT-D600 gives a larger AUC than CRAFT-4L, while CRAFT-6L achieves the best current AUC.

This result gives a practical scaling guideline. Width scaling is useful when increasing representation capacity at moderate depth. Depth scaling is useful when the model can preserve and refine intent information across more blocks. Under the current experimental budget, the strongest result comes from vertical scaling to six CRAFT blocks.

\subsection{Layer-wise Representation Dynamics}
\label{sec:representation_dynamics}
To examine how sequential information evolves across stacked blocks, we measure the effective rank of sequence-summary representations at each layer. Effective rank reflects how broadly representation energy is distributed across independent directions. We use it as a geometric diagnostic rather than a direct measure of representation quality.
\begin{figure}
    \centering
    \includegraphics[width=1.1\linewidth]{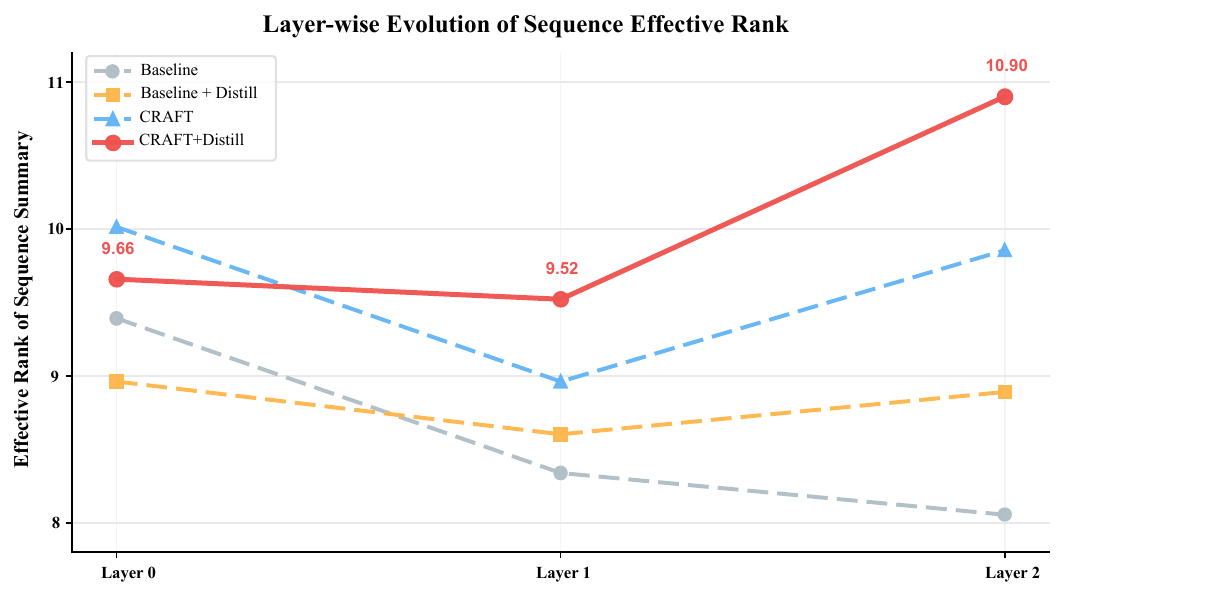}
    \caption{Layer-wise evolution of sequence effective rank. Layer 0 denotes the sequence summary before the first modeling block. The interaction baseline progressively compresses the sequence space, whereas CRAFT enables later blocks to recover task-relevant representation directions. Effective rank is used as a geometric diagnostic rather than a direct measure of representation quality.}
    \label{fig:seq_rank_evolution}
\end{figure}
As shown in Figure~\ref{fig:seq_rank_evolution}, the interaction baseline exhibits progressive compression, with its sequence rank decreasing from $9.393$ at Layer 0 to $8.057$ at Layer 2. Distillation alleviates this degradation, but its final rank ($8.892$) remains close to the initial value ($8.963$). CRAFT exhibits a different evolution pattern: its rank first decreases from $10.017$ to $8.964$, then recovers to $9.859$. This non-monotonic trajectory suggests that CRAFT does not indiscriminately expand representations. Instead, the first block filters sequence information, while subsequent contextual transport recovers task-relevant directions that are progressively lost by interaction-only updates.

Combining CRAFT with distillation produces the strongest effect. Its sequence rank remains stable in the first block ($9.661 \rightarrow 9.524$) and then increases to $10.903$, yielding a net expansion over the initial representation. Together with the AUC improvements in Table~\ref{tab:main_results}, these results indicate that CRAFT changes how sequential evidence is preserved and reorganized across depth, while distillation provides complementary predictive supervision. Therefore, the gain of distilled CRAFT cannot be attributed to distillation alone.

\section{Negative and Limited-Gain Findings}
\label{app:negative_findings}

Table~\ref{tab:appendix_negative} summarizes important negative and limited-gain explorations. These results are useful because they clarify the boundary of our method. The final improvement does not come from blindly adding complexity. Heavier dense tokenizers, larger FFN multipliers, and longer training do not automatically improve ranking quality. In some cases, they increase cost or instability.

The first group of findings concerns dense features. Structured dense vectors should not be flattened indiscriminately. When repeated segments, metadata suffixes, and stable vector prefixes are mixed by a generic projection, the model may learn unstable shortcuts. This weakens scaling because larger models can amplify these shortcuts. Structure-aware dense tokenization therefore acts as a reliability filter before feature transport.

The second group concerns capacity allocation. Increasing FFN expansion or adding heavier dense self-attention does not necessarily help. This suggests that recommendation scaling is not only about total parameter count. Capacity must be placed where the data structure supports it. In our experiments, useful capacity is better allocated to intent-mediated transport and carefully tokenized context representations.

The third group concerns training dynamics. Continuing training after the early best region often causes overfitting. This differs from many large-scale generative settings where long training can steadily improve performance. In CTR/CVR prediction, labels are noisy, data distribution is time-sensitive, and validation/test mismatch can appear quickly. Therefore, checkpoint selection, EMA, and early stopping are not minor engineering details; they are necessary for reliable model comparison.

Together, these negative findings support the final design. CRAFT works best when built on stable tokenization and training. The model should not transport raw heterogeneous noise more strongly; it should transport cleaned, structured, and reliability-aware context.

\section{Appendix Summary}
\label{app:summary}

The full system contains three coupled layers. First, heterogeneous raw features are converted into stable tokens through structure-aware tokenization and robust missing-value handling. Second, sparse and dense parameters are optimized with different training dynamics, and distillation is used as a separate training enhancement. Third, CRAFT performs contextual residual adaptive feature transport inside each unified block.

The evidence across ablation, representation diagnostics, and scaling analysis is consistent. CRAFT improves ranking performance without relying on a larger inference model. It increases intent and sequence representation diversity, organizes label-discrimin-ative information through intent carriers, and benefits from both depth and width scaling. The strongest current result is obtained by stacking CRAFT to six blocks, suggesting that controlled feature transport is a practical mechanism for scalable unified recommendation.

\end{document}